\documentclass[a4paper,10pt,twoside]{cpc-hepnp}

\usepackage{multicol}
\usepackage{graphicx}
\usepackage{booktabs}
\usepackage{amssymb,bm,mathrsfs,bbm,amscd}
\usepackage[tbtags]{amsmath}
\usepackage{lastpage}

\begin{document}

\fancyhead[c]{\small Chinese Physics C~~~Vol. xx, No. x (201x) xxxxxx}
\fancyfoot[C]{\small 010201-\thepage}


\title{Decay rates of charmonia within a quark-antiquark confining potential  }

\author{%
      Smruti Patel$^{1}$,\email{fizixsmriti31@gmail.com}%
\quad P. C. Vinodkumar$^{1},$\email{p.c.vinodkumar@gmail.com}%
\quad Shashank Bhatnagar$^{2}$\email{shashank-bhatnagar@yahoo.com}%
}
\maketitle

\address{%
$^1$ Department of Physics, Sardar Patel University,Vallabh Vidyanagar, INDIA.\\
$^2$ { Department of Physics, Addis Ababa University, P.O.Box 101739, Addis Ababa, Ethiopia.} \\
}

\begin{abstract}
In this work, we investigate the spectroscopy and decay
rates of charmonia within the framework of non-relativistic Schr\"{o}dinger equation by employing an approximate inter
quark-antiquark potential. The spin hyperfine, spin-orbit and tensor components of the one gluon exchange interaction are
employed to compute the spectroscopy of the excited S states and few low-lying P and D waves. The resultant wave functions at zero inter quark separation as well as some finite separation are employed to predict the di-gamma, di-leptonic and di-gluon decay rates of charmonia states by using the conventional Van Royen-Weisskopf formula. The di-gamma and di-leptonic decay widths are also computed by incorporating the relativistic corrections of order $v^4$ within the NRQCD formalism. We have observed that the NRQCD predictions with their matrix elements computed at finite radial separation yielded results which are found to be in better agreement with experimental value for both di-gamma and di-leptonic decays. The same scenario is seen in the case when di-gamma and di-leptonic decay widths are computed with Van Royen-Weisskopf formula. It is also observed that the di-gluon decay width with the inclusion of binding energy effects are in better agreement with the experimental data available for 1S-2S and 1P.  The di-gluon decay width of 3S and 2P waves waves are also predicted. Thus, the present study of decay rates clearly indicates the importance of binding energy effects.
\end{abstract}



\footnotetext[0]{\hspace*{-3mm}\raisebox{0.3ex}{$\scriptstyle\copyright$}2013
Chinese Physical Society and the Institute of High Energy Physics
of the Chinese Academy of Sciences and the Institute
of Modern Physics of the Chinese Academy of Sciences and IOP Publishing Ltd}%

\begin{multicols}{2}

\section{Introduction}

After a hiatus of about three decades, Charmonium has proved a remarkable laboratory for the study of quantum
chromodynamics(QCD). Following a period of several-years of intense experimental activity, charmonium physics has emerged
again as one of the most exciting areas of experimental high energy physics due to the massive dedicated investigations by
experimenters. A plenty of new data coming from the different experimental groups like BaBar, CLEO, SELEX, Tevatron and other B
factories world over and by the progress made in the theoretical methods in the last few years has greatly changed the thrust in this area \cite{G,Ecklund,Auger}. The study of quarkonium spectrum provides fundamental informations about the interquark potential. Yet, despite the apparent simplicity of these states, the mechanism behind their production remains a mystery, even after decades of experimental and theoretical efforts \cite{Nora}. The production rate in various high energy processes can give valuable insight into the heavy quark-antiquark interactions as well as into the elementary processes leading to the production of the ($c\bar c$) pair. Furthermore, these mesons enter a number of reactions which are of greatest importance for the study of the Cabibbo-Kobayashi-Maskawa(CKM) matrix and of CP violation.

The spectroscopy and decay rates of quarkonia are quite important to study as huge amount of high precession data
acquired from many experimental facilities world over are continuously providing accurate information about hadrons
particularly in charm and beauty flavour sectors \cite{Nakamura,olive}. Many theoretical predictions on the decay
properties particularly  the leptonic and di-gamma decays of quarkonia based on the relativistic quark model or
potential model \cite{Ahmady,Ebert,Hwang}, Bethe-Salpeter equation\cite{Huang,Kim}, heavy-quark spin symmetry \cite{JP}
and lattice QCD \cite{JJ} are available in literature.
The spectroscopic parameters like the interquark potential and its parameters that describe the masses of the low lying bound states and the resulting wave functions are important and decisive in the descriptions of other properties like the decay (in the annihilation channel) and transition rates. In the present study, we deduce the basic parameters of the mesonic states by fitting the masses of the low lying $c\bar c$ states based on a phenomenological potential framework.

The success of any theoretical model for mesons depends on the correct prediction of their decay rates apart from their mass spectra. In many phenomenological models the predictions of the masses are correct but prediction for the decay rates are overestimated \cite{WBuch,Amartin,CQ,Rai,SS}. The incorporation of various corrections due to radiative processes, higher-order QCD contributions etc. to decay rates have been suggested for better estimates of their decay properties with reference to the experimental data. In this context, the NRQCD formalism is found to provide a systematic treatment of the perturbative and non-perturbative components of QCD at the hadronic scale \cite{khan,GT,GT1,GT2}. For the present study, we employ phenomenological potential schemes for the bound states of heavy quarkonia and the resulting parameters and wave functions to study the decay properties. The study of di-gamma and di-lepton decay widths of charmonia has been done using the conventional Van Royen Weisskopt formula as well as using the NRQCD formalism.

The paper is organized as follows. In Section 2, we describe the phenomenological quark-antiquark interaction potential
and extract the parameters that describe the ground state masses of $c\bar c$ system. We also compute the low lying
orbital excited states of these systems. In Section 3 we employ the spectroscopic parameters of the $c\bar c$ system to
study the two photon and di-leptonic decay widths in conventional as well as NRQCD formalism. In Section 4 we present and analyze our results to draw important conclusions.

\section{The phenomenology and extraction of the spectroscopic parameters}

There are many methods to estimate the mass of a hadron, among which phenomenological potential model is a fairly reliable one, specially for heavy hadrons.
For the description of the quarkonium bound states, we adopt the phenomenological potential of the form which is expressed
in terms of a vector (Coulomb) plus a scalar (confining) part given by
\begin{equation}\label{eq:1}
V(r)=V_V+V_S=\frac{-4\alpha_s}{3r}+\frac{Ar^2}{({1+4Br^n})^{\frac{1}{2}}}-V_0
\end{equation}

Here, A=0.374 $GeV^3$, B=1.0 $GeV^n$ and $V_s$ is a state dependant constant potential. Here, $\alpha_s$ is the running
strong coupling constant which is computed as,
\begin{equation}\label{eq:2}
\alpha_s(\mu^2)=\frac{4\pi}{(11-\frac{2}{3}n_f)(\ln\frac{\mu^2}{\Lambda^2})}
\end{equation}

where, $\Lambda$ is the QCD scale which is taken as 0.120 $GeV$, $n_f$ is the number of flavors, $\mu$ is the renormalization scale related to the constituent quark mass. Similar type of potential with $n=2$ has been used by \cite{ANM,Shashank,SB} for the study mainly of ground state light flavor hadrons using the field theoretic framework of Bethe-Salpeter equation under Covariant Instantaneous Ansatz(CIA), which is a Lorentz-invariant generalization of Instantaneous Ansatz (IA). Such type of potential in the above framework  was employed for calculations \cite{Shashank,SB} dealing with studies on leptonic decay constants of ground state vector mesons ($\rho$, $\phi$, $\omega$, $\psi$, $Y$) as well as ground state pseudoscalar mesons ($\pi$,$K$, $D$, $D_s$, $B$), two-photon decay widths for the process, $P\rightarrow\gamma \gamma$ and radiative decay widths of light vector mesons through the process $V\rightarrow P\gamma$. In all these studies the confining term in the potential in Eq. (1) is supposed to simulate an effect of an almost linear confinement $(~ r)$ for heavy quark (c, b) sector, while retaining harmonic form $(~ r^2)$ for light quark(u,d) sector as is believed to be true for QCD.

Further, in these studies, the main ingredient is the 4 dimensional(4D) hadron-quark vertex function, which plays the role of an exact effective coupling vertex of the hadron with all its constituents (quarks). The 4D BS wave function (comprising of hadron-quark vertex) is considered to sum up all the non-perturbative QCD effects in the hadron. The hadron-quark vertex has been employed for calculation of transition amplitudes for the above mentioned processes through quark-loop diagrams, with parameters fixed from the mass spectra of mesons, which was obtained by solving a 3 dimensional Salpeter equation derived from the reduction of the full 4D BSE \cite{SB}. Now one of the main ingredients in 4D BS wave function (BSW) is its Dirac structure.
Recent studies \cite{alkofer,alkofer1,cvetic} have revealed that various mesons have many different Dirac structures in their BS wave functions, whose inclusion is necessary to obtain quantitatively accurate observables and all structures do not contribute equally for calculation of various meson observables. Also, many hadronic processes are particularly sensitive to higher order Dirac structures in BS amplitudes. Towards this end, to ensure a systematic procedure of incorporating various Dirac covariants
from their complete set in the BS wave functions of various hadrons (pseudoscalar, vector etc.), a naive power counting rule
was developed in \cite{Shashank,SB}, to enable one to identify the leading Dirac structures from the sub-leading ones, and to study the relevance of various Dirac structures in calculation of different meson observables in these calculations. The potential with $n=2$ used in Eq. (1) was shown to give quite accurate results for a vast range of processes in the framework of BSE, with parameters fixed from the ground state mass spectrum. Further, the calculation of decay widths of strong decays of radially excited vector meson states through the process $V'\rightarrow PP$ in BSE\cite{elias} was also recently attempted for the light sector only, using the leading Dirac covariants alone, employing the above form of potential of Eq. (1). So present study is mainly an extension of such type of potentials in the quarkonia sector by computing the spectra for radial as well orbital excitations and also by calculating decay widths for a vast range of processes, since the success of any potential depends on the depth and range of its predictions for a vast range of processes.

The potential with index $n=2$ is found to be inconsistent in correctly predicting the hyperfine splitting between pseudoscalar and vector mesons even for the ground state. Besides for orbital excited states the potential behavior becomes repulsive with the use of $r^2$ term in the denominator. While with the choice of $n=1$ in the Eq. (1), for the potential the overall shape of the potential (see inserted figure at right bottom corner of fig 1) is not altered much and found better consistancy for the predictions of hyperfine and fine structure splitting of the $c\bar c$ states. Moreover, the potential with power $n=1$ is more shallower than the potential with $n=2$ and this shallow nature is required for the excited state predictions of heavy quarkonia.

We now present the details of our calculations by using the potential with $n=1$ of Eq. (1). Different degenerate n$^{2S+1}L_{J}$ low-lying states of $c\bar c$ mesons are calculated by including spin dependent part of the usual one gluon exchange potential \cite{Barnes,Olga,Voloshin,Eichten,SS}. The potential description extended to spin dependent interactions results in three types of potential terms such as the spin-spin, the spin-orbit and the tensor part that are to be added to the spin independent potential as given by Eq. (1). Accordingly, the spin-dependent part $V_{SD}$ is given by

\begin{eqnarray}\label{eq:3}
V_{SD}&=&V_{SS}\left[\frac{1}{2}(S(S+1)-\frac{3}{2}))\right] \nonumber \\ && +V_{LS}\left[\frac{1}{2}(J(J+1)-S(S+1)-L(L+1))\right] \nonumber \\ && +V_{T}\left[12\left(\frac{(S_1.r)(S_2.r)}{r^2}-\frac{1}{3}(S_1.S_2)\right)\right]
\end{eqnarray}

The spin-orbit term containing $V_{LS}$ and tensor term containing $V_T$ describe the fine structure of the states, while the spin-spin term containing $V_{SS}$ proportional to 2$S_1.S_2$ gives the hyperfine splitting. The co-efficient of these spin-dependent terms of Eq.(\ref{eq:10}) can be written in terms of the vector and scalar parts of static potential
V(r) as

\begin{equation}\label{eq:4}
V^{ij}_{LS}(r)=\frac{1}{2M_iM_jr}\left[ 3\frac{dV_V}{dr}-\frac{dV_S}{dr}\right]
\end{equation}

\begin{equation}\label{eq:5}
V^{ij}_T(r)=\frac{1}{6M_iM_j}\left[ 3\frac{d^2V_V}{dr^2}-\frac{1}{r}\frac{dV_S}{dr}\right]
\end{equation}

\begin{equation}\label{eq:6}
V^{ij}_{SS}(r)=\frac{1}{3M_iM_j}\nabla^2 V_V=\frac{16\pi\alpha_s}{9M_iM_j}\delta^3(r)
\end{equation}

Where $M_i$, $M_j$ corresponds to the quark masses. The Schr\"{o}dinger equation with the potential given by
Eq.(\ref{eq:1}) is numerically solved using the Mathematica notebook of the Runge-Kutta method \cite{Lucha} to
obtain the energy eigen values and the corresponding wave functions.

The computed masses of the nS, nP and nD states are listed in Table I, II and III respectively. The optimized
spectroscopic parameters thus correspond to the fitted quark masses, the potential strength $V_0$ and the corresponding
radial wave functions. The quark mass $m_c$ = 1.28 GeV, while the potential strength ($V_0$ ) is given by the relation
\begin{eqnarray}\label{eq:7}
V_0(n+1,l)=V_0(n)+0.02l(3l+5)+\frac{1}{2}
\end{eqnarray}

Where $l$ is the orbital angular momentum. The value of $V_0$(n = 0; l = 0) is fixed as 0.12 $GeV$. We have plotted the
behaviour of the present potential for different states and are shown in figure 1.
\begin{figure}
\resizebox{0.47\textwidth}{!}{%
\includegraphics{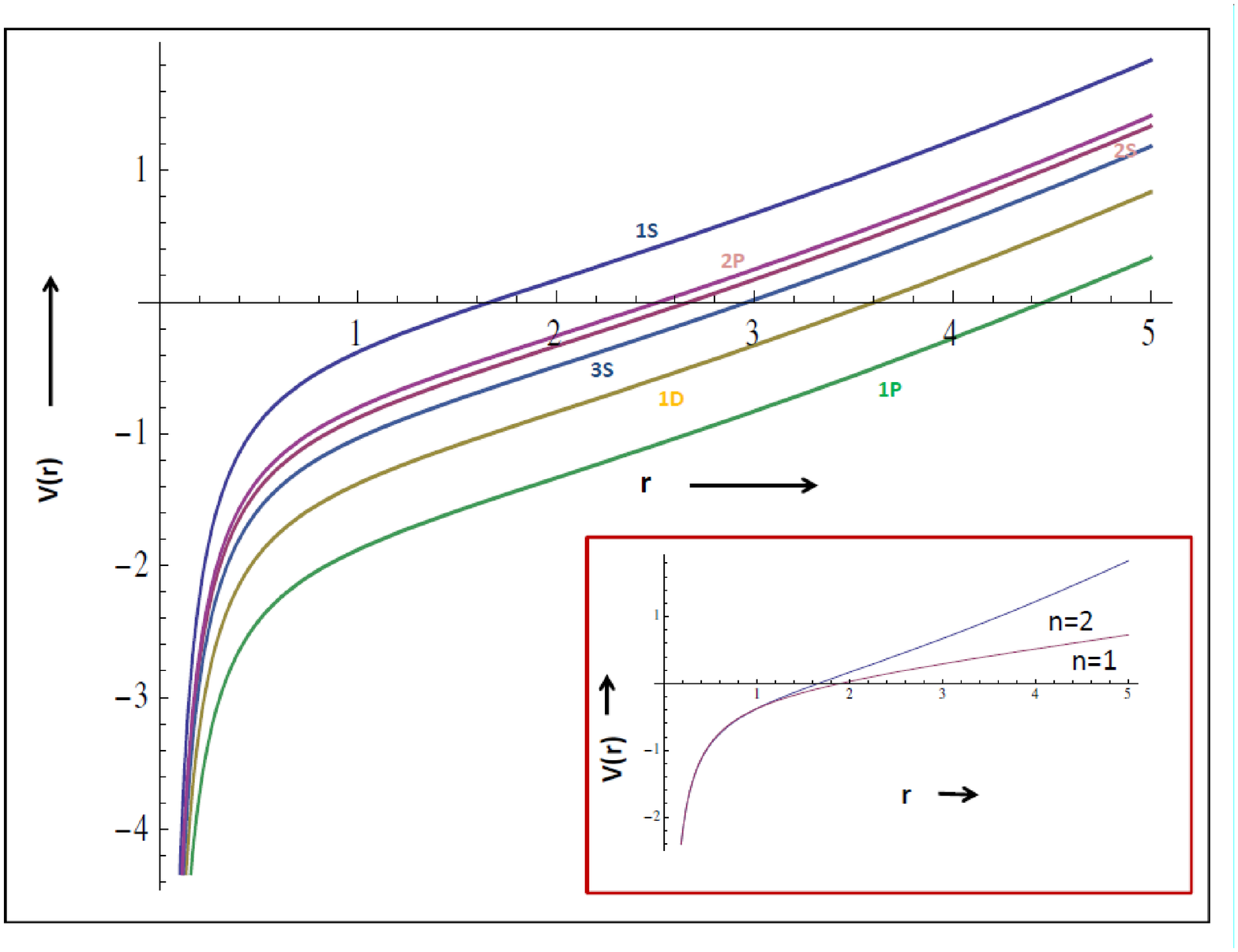}
}
\caption{Behaviour of the potential V(r) at quark-antiquark separation(r)}\label{modified}

\end{figure}




\section{Decay rates of heavy quarkonia}

Apart from the masses of the low lying mesonic states, the correct predictions of the decay rates are important features of any
successful model. There have been a number of recent studies on processes involving strong decays, radiative decays and leptonic decays of vector mesons. Such studies offer a direct probe of hadron structure and help in revealing some aspects of the
underlying quark-gluon dynamics that are complementary to what is learnt from pseudo scalar mesons. Leptonic decay constants are simple probes of the short distance structure of hadrons and therefore are a useful observable for testing quark dynamics in
this regime. The extracted model parameters and the radial wave functions are being employed here to compute the di-leptonic,
two-photon and two gluon annihilation rates and since this rate is related to the wave function, it provides better understanding of the quark-antiquark dynamics within the meson. It can be a crucial test for potential model. The radiative decays of the bound $c\bar c$ states provide an excellent laboratory for studying charmonium decay dynamics and the light hadron spectroscopy. An electromagnetic decay occurs when the $c\bar c$ pair annihilates into one or more photons, which can subsequently lead to a pair of leptons as the final state. These processes can be calculated with perturbative quantum electrodynamics (QED) with corrections from the strong interaction.

\subsection{Using Van Royen-Weisskopf formula}

A decay to a pair of leptons is only allowed to the states with the same quantum numbers as the photon, that is $J^{PC}=1^{--}$. Using the Van Royen-Weisskopf formula the leptonic decay width with radiative correction for the vector mesons reads:

\begin{equation}\label{eq:8}
\Gamma({n^3S_{1}\longrightarrow}{l^+l^-})=\frac{4N_c\alpha^2 {e_Q}^2|R_{nl}(r)|^2}{M{_V}^2}\left[1-\frac{16}{3}\left(\frac{\alpha_S}{\pi}\right)\right]
\end{equation}

A decay into two photons is instead forbidden to the $J = 1$ states by the Yang theorem \cite{Landau,Yang}.
For other resonances, the conservation of charge parity requires the S wave states to be in a spin-singlet
state and the P-wave states to be in a spin-triplet state. For the S wave di-gamma decay widths, most of the model
predictions are consistent with experimental results, while in the case of P waves the theoretical predictions for
digamma widths differs from the experimental results. This discrepancy is somehow removed with the inclusion of QCD
corrections. The di-gamma decay widths for the $\eta_Q$, $\chi_{Qo}$, $\chi_{Q2}$ into two photons with one loop
radiative corrections are computed using the non-relativistic expression given by \cite{Kwong,Parmar,BKP,ajay,Rai,HWH,RB,AP}

\begin{eqnarray}\label{eq:9}
\Gamma({n^1S_{0}\longrightarrow{\gamma\gamma}})&=&\frac{3}{2}\frac{\alpha^2 {e_Q}^4 M_{\eta_Q}|R_{nl}(r)|^2}{M{_Q}^3}\times \nonumber \\&&
\left[1-\frac{(20-{\pi}^2)}{3}\left(\frac{\alpha_S}{\pi}\right)\right]
\end{eqnarray}

\begin{eqnarray}\label{eq:10}
\Gamma({n^3P_{0}\longrightarrow{\gamma\gamma}})&=&\frac{27\alpha^2 {e_Q}^4 M_{\chi_{Q0}}|R^{(l)}_{nl}(r)|^2}{2M{_Q}^3}\times \nonumber \\&&
\left[1+\frac{({\pi}^3)}{3}\frac{28}{9}\left(\frac{\alpha_S}{\pi}\right)\right]
\end{eqnarray}

\begin{eqnarray}\label{eq:11}
\Gamma({n^3P_{2}\longrightarrow{\gamma\gamma}})&=&\frac{4}{15}\frac{27\alpha^2 {e_Q}^4 M_{\chi_{Q2}}|R^{(l)}_{nl}(r)|^2}{2M{_Q}^3}\times \nonumber \\&&
\left[1-\frac{16}{3}\left(\frac{\alpha_S}{\pi}\right)\right]
\end{eqnarray}

Among hadronic decays, we can consider annihilations and transitions. The first type of decays occur when the $c\bar c$
pair annihilates into two or more gluons or light quarks. In analogy to the electromagnetic decays, a decay into two gluons
$gg$ is allowed to the same states which can decay into, with respect to which it is much more favoured due to the larger
coupling constant. The di-gluon decay width gives information on the total width of the corresponding quarkonium \cite{Wang}.

The relevant theoretical expressions for the di-gluon decay widths of n$^1S_0$, n$^3P_0$ and n$^3P_2$ charmonia states,
incorporating the leading order QCD corrections, are given by \cite{Kwong,Lansberg,Barbieri,Petrelli}

\begin{equation}\label{eq:12}
\Gamma_{gg}(\eta_Q)=\frac{\alpha_s^2M_{\eta_Q}|R^{(l)}_{nl}(r)|^2}{3m{_Q}^3}\left[ 1+4.8\left(\frac{\alpha_S}{\pi}\right)\right]
\end{equation}

\begin{equation}\label{eq:13}
\Gamma_{gg}({\chi_{Q_0}})=\frac{3\alpha_s^2M_{\chi_{Q_{c0}}}|R^{(l)}_{nl}(r)|^2}{m{_Q}^5}\left[ 1+8.77\left(\frac{\alpha_S}{\pi}\right)\right]
\end{equation}

\begin{equation}\label{eq:14}
\Gamma_{gg}({\chi_{Q_2}})=\frac{4}{15}\frac{3\alpha_s^2M_{\chi_{Q_{c2}}}|R^{(l)}_{nl}(r)|^2}{m{_Q}^5}\left[1-4.827\left(\frac{\alpha_S}{\pi}\right)\right]
\end{equation}

Here, $\alpha  = 1/137$ is the electromagnetic coupling constant and ${e_Q}$ corresponds to the charge content of the $Q\bar Q$ meson in terms of the electron charge. For $c\bar c$ meson, ${e_Q}=2/3$ and $m_Q=m_c$.
Within the potential confinement scheme, we consider the constituent quark mass $m_c$ appeared in Eqs.(\ref{eq:9}-\ref{eq:14}) as effective mass of the quark within the bound state of the charmonium system as defined as \cite{BP,bhavin}
\begin{equation}\label{eq:15}
m^{eff}_c=m_c\left(1+\frac{{\langle E_{bind}\rangle}_{nl}}{m_c+m_{\bar c}}\right)
\end{equation}

\begin{table*}
\begin{center}
\caption{Charmonium mass spectra for nS states in GeV}. \label{tab2}
\begin{tabular}{cc|cccccccc}
\hline
&   State   &   Present &   \cite{Bali} &   \cite{olive}    &   \cite{Lakhina}  &   \cite{BKP}  &   \cite{Barnes}   &    \cite{OM}  &    \cite{MNS} \\\hline
&   $1^3S_1$    &   3.096   &   3.175   &   3.097   &   3.168   &   3.090   &   3.090   &   $3.085\pm0.001$ &    3.097  \\
&   $1^1S_0$    &   2.979   &   2.966   &   2.980   &   3.088   &   2.976   &   2.982   &   $3.010\pm0.001$ &    2.980  \\
&    $2^3S_1$   &   3.680   &   3.705   &   3.686   &   3.707   &   3.615   &   3.672   &   $3.739\pm0.046$ &    3.687  \\
&   $2^1S_0$    &   3.600   &   3.560   &   3.638   &   3.669   &   3.533   &   3.630   &   $3.770\pm0.040$ &    3.631  \\
&    $3^3S_1$   &   4.077   &   4.106   &   4.040   &   4.094   &   3.962   &   4.072   &   -   &   4.030   \\
&   $3^1S_0$    &   4.011   &   3.978   &   -   &   4.067   &   3.895   &   4.063   &   -   &   3.992   \\
&   $4^3S_1$    &   4.454   &   4.442   &   4.415   &   4.420   &   4.240   &   4.406   &   -   &   4.273   \\
&   $4^1S_0$    &   4.397   &   4.324   &   -   &   4.398   &   4.180   &   4.384   &   -   &   4.244   \\

\hline

\end{tabular}

\begin{center}
\cite{olive}-Exp\\
\cite{Bali,OM}-Lattice\\
\cite{Lakhina}-NRQM\\
\cite{Barnes,BKP,MNS}-Potential models
\end{center}
\end{center}

\end{table*}
\begin{table*}
\begin{center}
\caption{Charmonia spectra for nP(L=1,2) waves in GeV}. \label{tab1}
\begin{tabular}{ccccccccccccc}
\hline
state   &   Mcw &   $n^{2S+1}L_J$&  $V_T$   &   $V_{LS}$    &   Present&    Exp.\cite{olive}    &   \cite{Bali} &   \cite{Lakhina}& \cite{BKP}  &   \cite{Barnes}    &   \cite{OM}   \\
&&&contribution&contribution&&&&&&&\\
\hline
1P  &   3.539   &   $1^3P_2$    &   -0.000006   &   0.025   &   3.565   &   3.556   &   3.491   &   3.564   &   3.524   &   3.556   &   $3.503\pm0.024$ \\
    &       &   $1^3P_1$    &   0.00003 &   -0.025  &   3.514   &   3.510   &   3.490   &   3.520   &   3.514   &   3.505   &   $3.472\pm0.009$ \\
    &       &   $1^3P_0$    &   -0.00006    &   -0.05   &   3.488   &   3.414   &   3.442   &   3.448   &3.466  &   3.424   &   $3.408\pm0.002$ \\
&       &   $1^1P_1$    &       &       &   3.539   &   3.526   &   3.486   &   3.536   &3.514  &   3.516&$3.474\pm0.010$   \\
&       &       &       &       &       &       &       &       &       &       &       \\
2P  &   3.996   &   $2^3P_2$    &   -0.000004   &   0.0247  &   4.021   &   3.929   &   3.924   &   -   &   -   &-  &$4.030\pm0.180$    \\
&       &   $2^3P_1$    &   0.000018    &   -0.0247 &   3.972   &   -   &   3.917   &   -   &   -   &   -   &   $4.067\pm0.105$ \\
&       &   $2^3P_0$    &   -0.000037   &   -0.0495 &   3.947   &   -   &   3.870   &   -   &   -   &   -   &   $4.008\pm0.122$ \\
    &       &   $2^1P_1$&-&-&3.996&-&3.916&-&-&-&$4.053\pm0.095$    \\

\hline

\end{tabular}
\end{center}
\end{table*}
\begin{table*}
\begin{center}
\caption{Charmonia spectra for nD(n=1,2) waves in GeV}. \label{tab1}
\begin{tabular}{cccccccccccc}
\hline
state   &   Mcw &   $n^{2S+1}L_J$&  $V_T$   &   $V_{LS}$    &   Present &   Exp.\cite{olive}    &   \cite{Bali} &   \cite{Lakhina}& \cite{BKP}  &   \cite{Barnes}        \\
&&&contribution&contribution&&&&&&\\
\hline
1D  &   3.796   &   $1^3D_3$    &    0.0023  &   -0.000008  &   3.798   &       &   3.770   &   3.809   &   3.83    &   4.167   &   \\
    &       &   $1^3D_2$    &-0.0011    &   0.00003 &   3.794   &       &   3.792   &   3.804   &   3.854   &   4.158   &   \\
    &       &   $1^3D_1$    &    -0.0034   &   -0.00003 &   3.792   &   3.770   &   3.796   &   3.789   &   3.860   &   4.142   &   \\
    &       &   $1^1D_2$    &       &       &   3.796   &       &   3.782   &   3.803   &   3.844   &   4.158   &   \\
    &       &       &       &       &       &       &       &       &       &       &   \\
2D  &4.224  &   $2^3D_3$    & 0.0012   &     -0.000003  &   4.425   &       &       &       &       &       &   \\
    &       &   $2^3D_2$    &-0.0006 &      0.00001 &   4.223   &       &       &       &       &       &   \\
    &       &   $2^3D_1$    &  -0.0019   &    -0.00001  &   4.222   &   4.160   &       &       &       &       &   \\
    &       &   $2^1D_2$    &       &       &   4.224   &       &       &       &       &       &   \\

\hline

\end{tabular}

\end{center}

\end{table*}
\begin{table*}
\begin{center}
\caption{Charmonium mass splitting compared to experimental and other predictions(in MeV))}. \label{tab1}
\begin{tabular}{ccccccc}
\hline
Mass difference &   Present &   \cite{daniel}   &   \cite{MNS}  &   \cite{ajay} &   Experiment  \\
\hline
    &       &       &       &       &       \\
1P-1S splitting &   471 &   $457.3\pm3.6$   &   455 &   863.5   &   $457.5\pm0.3$   \\
1S hyperfine    &   118 &   $118.1\pm2.1^{-1.5}_{-4.0}$ &   116.74  &   174 &   $113.2\pm0.7$   \\
1P spin-orbit   &   34.11   &   $49.5\pm2.5$    &   65.88   &   -   &   $46.6\pm0.1$    \\
1P tensor   &   20.11   &   $17.3\pm2.9$    &   13.17   &   -   &   $16.25\pm0.22$  \\
2S-1S splitting &   593 &   -   &   606 &   529 &   $606\pm1$   \\
\hline

\end{tabular}

\end{center}

\end{table*}

\begin{table*}
\begin{center}
\caption{Di-leptonic decay widths of charmonium in keV}. \label{tab1}
\begin{tabular}{cc|cccc|ccc}
\hline
&   State   &   \ \ \ \ \ \ \ \ Present         &           && &\cite{olive} &  \cite{MNS}  &\cite{Stanley} \\ \hline
&       &   $\Gamma_0(0)$&  $\Gamma_{0R}(0)$    &$\Gamma_0(r_0)$    &   $\Gamma_{0R}(r_0)$          &       &   \\\hline
&   $J/\psi(1S)$    &   9.22    &   4.61    &   5.01    &   2.50    &   $5.55\pm0.14$ &4.95 &1.85   \\
&   $\psi(2S)$  &   6.87    &   3.43    &2.33  &    1.16    &   $2.35\pm0.04$   &   2.33& 0.89    \\
&   $\psi(3S)$  &   5.89    &   3.04    &   1.64        &   0.820   &   $0.86\pm0.004$& 1.63&   0.98\\

\hline

\end{tabular}

\end{center}

\end{table*}
\begin{table*}
\begin{center}
\caption{Di-gamma decay widths of charmonia states(S and P waves)using $m_{eff}$ in keV}. \label{tab1}
\begin{tabular}{c|ccccccc}
\hline

state   &   $1^1S_0$    &   $2^1S_0$    &   $1^3P_0$    &   $1^3P_2$    &   $2^3P_0$    &   $2^3P_2$    \\
\hline
$\Gamma_{0}(0)$ &   11.49   &   8.873   &   9.964   &   1.358   &   5.46    &   1.48    \\
$\Gamma_{0R}(0)$    &   7.853   &   6.061   &   5.065   &   0.68    &   2.97    &   0.74    \\
$\Gamma_{0}(r_0)$   &   4.022   &   1.869   &   8.789   &   1.197   &   4.02    &   1.09    \\
$\Gamma_{0R}$  &   2.747   &   1.276   &   4.468   &   0.599   &   1.04    &   0.54    \\
    &       &       &       &       &   &       \\
Others  &   $5.055\pm0.411^{*}$\cite{olive} &   $2.147\pm1.580$\cite{olive} &   $2.341\pm0.189$ \cite{olive}&   $0.528\pm0.404$ \cite{olive}&   $1.7$\cite{Bai} &    $0.23$\cite{Bai}   \\
    &   10.37\cite{MNS} &   3.349\cite{MNS} &   2.5\cite{Bai}   &   0.31\cite{Bai}  &       &       \\
    &   8.5\cite{Bai}   &   2.4\cite{Bai}   &   $2.36\pm0.35$\cite{CW}  &   $0.346^{+0.009}_{0.011}$\cite{CW}   &       &       \\
    &       &       &   5.0\cite{Lansberg}  &   0.70\cite{Lansberg}     &       &       \\
    &       &       &   6.38\cite{SNG}  &   0.57\cite{SNG}  &   &       \\
    &       &       &   3.96\cite{crater}   &   0.743\cite{crater}  &       &   \\

\hline
\end{tabular}
\begin{center}
*The di-gamma decay widths are estimated using the values of branching fraction and full decay width given in PDG[2014]
\end{center}
\end{center}

\end{table*}
\begin{table*}
\begin{center}
\caption{Di-gluon decay for nS states for charmonia states in MeV}. \label{tab1}
\begin{tabular}{c|cccc}
\hline
State       &   Decay width &   Present &   Others  \\
\hline
    &   $\Gamma^{m}_{gg}(0)$    &   38.47   &   32.20\cite{AP}  \\
    &   $\Gamma^{m}_{ggR}(0)$   &   55.80   &   10.70\cite{james}   \\
    &   $\Gamma^{meff}_{gg}(0)$ &   22.37   &   19.60\cite{Kim} \\
$1^1S_0$    &$\Gamma^{meff}_{ggR}(0)$   &   32.45   &   23.03\cite{SNG} \\
    &   $\Gamma^{m}_{gg}(r_0)$  &   12.70   &   9.010\cite{MGO} \\
    &   $\Gamma^{m}_{ggR}(r_0)$ &   18.42   &   $26.7\pm3.0$\cite{olive}    \\
    &   $\Gamma^{meff}_{gg}(r_0)$   &   7.38    &       \\
    &   $\Gamma^{meff}_{ggR}(r_0)$  &   10.71   &       \\
    &                       \\
    &   $\Gamma^{m}_{gg}(0)$    &   47.47   &   8.10\cite{james}    \\
    &   $\Gamma^{m}_{ggR}(0)$   &   38.86   &   12.1\cite{Kim}  \\
    &   $\Gamma^{meff}_{gg}(0)$ &   16.74   &   $14.7\pm0.7$\cite{olive}    \\
$2^1S_0$    &   $\Gamma^{meff}_{ggR}(0)$    &   24.29   &       \\
    &   $\Gamma^{m}_{gg}(r_0)$  &   10.02   &       \\
    &   $\Gamma^{m}_{ggR}(r_0)$ &   14.54   &       \\
    &   $\Gamma^{meff}_{gg}(r_0)$   &   3.558   &       \\
    &   $\Gamma^{meff}_{ggR}(r_0)$  &   5.133   &       \\
    &                       \\
    &   $\Gamma^{m}_{gg}(0)$    &   56.02   &       \\
    &   $\Gamma^{m}_{ggR}(0)$   &   81.26   &       \\
    &   $\Gamma^{meff}_{gg}(0)$ &   14.03   &       \\
$3^1S_0$&   $\Gamma^{meff}_{ggR}(0)$    &   20.36   &       \\
    &   $\Gamma^{m}_{gg}(r_0)$  &   7.170   &       \\
    &   $\Gamma^{m}_{ggR}(r_0)$ &   10.40   &       \\
    &   $\Gamma^{meff}_{gg}(r_0)$   &   1.796   &       \\
    &   $\Gamma^{meff}_{ggR}(r_0)$  &   2.606   &       \\
\hline

\end{tabular}

\end{center}

\end{table*}
\begin{table*}
\begin{center}
\caption{Di-gluon decay for P waves for charmonia states in MeV}. \label{tab1}
\begin{tabular}{c|cccc}
\hline
State       &   Decay width &   Present &   Others  \\
\hline
    &   $\Gamma^{m}_{gg}(0)$    &   47.88   &   10.46\cite{AP}  \\
    &   $\Gamma^{m}_{ggR}(0)$   &   81.18   &   13.44\cite{SNG} \\
    &   $\Gamma^{meff}_{gg}(0)$ &   9.45    &   $12.5\pm3.2$\cite{HHW}  \\
$1^1P_0$    &$\Gamma^{meff}_{ggR}(0)$   &   17.21   &   $10.4\pm0.7$\cite{olive}    \\
    &   $\Gamma^{m}_{gg}(r_0)$  &   41.07   &       \\
    &   $\Gamma^{m}_{ggR}(r_0)$ &   74.77   &       \\
    &   $\Gamma^{meff}_{gg}(r_0)$   &   8.271   &       \\
    &   $\Gamma^{meff}_{ggR}(r_0)$  &   15.05   &       \\
    &                       \\
    &   $\Gamma^{m}_{gg}(0)$    &   14.02   &   1.169\cite{AP}  \\
    &   $\Gamma^{m}_{ggR}(0)$   &   7.82    &   1.2\cite{SNG}       \\
    &   $\Gamma^{meff}_{gg}(0)$ &   2.81    &       1.72\cite{james}    \\
$1^3P_2$    &   $\Gamma^{meff}_{ggR}(0)$    &   1.54    &$2.03\pm0.12$\cite{olive}      \\
    &   $\Gamma^{m}_{gg}(r_0)$  &   12.62   &       \\
    &   $\Gamma^{m}_{ggR}(r_0)$ &   6.922   &   \\
    &   $\Gamma^{meff}_{gg}(r_0)$   &   2.465&  \\
    &   $\Gamma^{meff}_{ggR}(r_0)$  &   1.351   &       \\

\\
&   $\Gamma^{m}_{gg}(0)$    &   103.2   &   9.61\cite{Wang} \\
    &   $\Gamma^{m}_{ggR}(0)$   &   187.9   &       \\
    &   $\Gamma^{meff}_{gg}(0)$ &   10.09   &       \\
$2^1P_0$    &$\Gamma^{meff}_{ggR}(0)$   &   18.38   &       \\
    &   $\Gamma^{m}_{gg}(r_0)$  &   27.69   &       \\
    &   $\Gamma^{m}_{ggR}(r_0)$ &   15.18   &       \\
    &   $\Gamma^{meff}_{gg}(r_0)$   &   2.70    &       \\
    &   $\Gamma^{meff}_{ggR}(r_0)$  &   1.48    &       \\
    &                       \\
    &   $\Gamma^{m}_{gg}(0)$    &   75.06   &       \\
    &   $\Gamma^{m}_{ggR}(0)$   &   136.6   &           \\
    &   $\Gamma^{meff}_{gg}(0)$ &   7.34    &           \\
$2^3P_2$    &   $\Gamma^{meff}_{ggR}(0)$    &   13.37   &       \\
    &   $\Gamma^{m}_{gg}(r_0)$  &   20.39   &       \\
    &   $\Gamma^{m}_{ggR}(r_0)$ &   11.18   &   \\
    &   $\Gamma^{meff}_{gg}(r_0)$   &   1.99&   \\
    &   $\Gamma^{meff}_{ggR}(r_0)$  &   1.09    &       \\

\hline

\end{tabular}

\end{center}

\end{table*}
\subsection{Using NRQCD formalism}
\begin{table*}
\begin{center}
\caption{Di-leptonic decay widths of charmonia states using NRQCD formalism in keV}. \label{tab1}
\begin{tabular}{c|cccc}
\hline

state   &   $1^3S_1$    &   $2^3S_1$    &   $3^3S_1$     &      \\
\hline\hline
$\Gamma^{NRQCD}_0(0)$ &   7.280   &   3.397   &   3.089   &           \\
$\Gamma^{NRQCD}_{0R}(0)$    &   7.190   &   3.354   &   3.25   &       \\
$\Gamma^{NRQCD}({r_0})$   &   6.730   &   1.218   &   0.82   &       \\

Others  &   $5.55\pm0.14$\cite{olive}     &   $2.35\pm0.04$\cite{olive} &   $0.86\pm0.004$ \cite{olive}&   \\
    &   2.809\cite{BT} &   2.14\cite{bannur} &   0.796\cite{bannur}   &         \\
    &   4.698\cite{Amartin}   &   1.22\cite{Gonz}  &   0.76\cite{Gonz}  &          \\
    &     10.294\cite{EE}  &       &     &          \\
    &     5.470\cite{bannur}  &       &     &          \\
    &     2.94\cite{Gonz}  &       &     &          \\
\hline
\end{tabular}

\end{center}

\end{table*}
\begin{table*}
\begin{center}
\caption{Di-gamma decay widths of charmonia states using NRQCD formalism in keV}. \label{tab1}
\begin{tabular}{c|cccc}
\hline

state   &   $1^1S_0$    &   $2^1S_0$    &   $3^1S_0$     &      \\
\hline\hline \\
$\Gamma^{NRQCD}_0(0)$  &  18.19   & 14.01     &  13.93     &           \\
$\Gamma^{NRQCD}_{0R}(0)$ & 13.30   & 10.34    &   8.81  &          \\
$\Gamma^{NRQCD}({r_0})$   &  6.91   &   3.27   &   1.94   &       \\

Others  &   $5.055\pm0.411$\cite{olive}  &   $2.147\pm1.580$\cite{olive}    &   $2.341\pm0.189$ \cite{olive}&        \\
        &   6.561\cite{BT}               &   $4.44\pm0.48$\cite{Kim}              &  1.21\cite{Lakhina}           &   \\
        &    10.691\cite{Amartin}        &   1.8\cite{Ebert}                       &                               &         \\
        &  17.447\cite{EE}               &   3.5-4.5\cite{Lansberg}                   &                               &          \\

\hline
\end{tabular}

\end{center}

\end{table*}

The new role of the heavy flavour studies as the testing ground for the non-perturbative aspects of QCD, demands extension of earlier phenomenological potential model studies on quarkonium masses to their predictions of decay widths with the non-perturbative approaches like NRQCD. It is expected that the NRQCD formalism has all the corrective contributions for the right predictions of the decay rates. The decay rates of the heavy-quarkonium states into photons and pairs of leptons are among the earliest applications of perturbative quantum chromodynamics (QCD)\cite{RB,Appel}. In NRQCD formalism decay rates are factorized into short and long distance parts. The short-distance factor is related to the annihilation rate of the heavy quark and antiquark and this part is calculated in terms of the running coupling constant $\alpha_s (m_{Q})$ of QCD, evaluated at the scale of the heavy-quark $m_{Q}$, while the long-distance factor which contains all nonperturbative effects of the QCD is expressed in terms of the meson's nonrelativistic wave function or derivatives of wavefunctions, evaluated at origin.
Our attempt in this section is to study the di-gamma and di-lepton decay widths based on the NRQCD formalism \cite{GT}. NRQCD factorization expressions for the decay widths of quarkonia  are given by \cite {ajay,bodwin}\\ \\

\begin{eqnarray}\label{eq:nq1}
\nonumber && \Gamma(^1 S_0 \rightarrow \gamma \gamma)  =
\frac{F_{\gamma
\gamma}(^1S_0)}{m^2_Q} \left |\left<0|\chi^{\dag}\psi|^1 S_0\right>\right|^2 \\
 \nonumber  && +
\frac{G_{\gamma \gamma}(^1S_0)}{m^4_Q} Re \left
[\left<^1S_0|\psi^{\dag}\chi|0\right>\left<0|\chi^{\dag}\left(-\frac{i}{2}\overrightarrow{D}\right)^2\psi|^1S_0\right>\right]
\cr \\
\nonumber   &&+ \frac{H^1_{\gamma \gamma}(^1S_0)}{m^6_Q}\left<^1
S_0|\psi^{\dag}\left(-\frac{i}{2}\overrightarrow{D}\right)^2\chi|0\right> \times\\
&& \left<0|\chi^{\dag}\left(-\frac{i}{2}\overrightarrow{D}\right)^2\psi|^1S_0\right> + \frac{H^2_{\gamma \gamma}(^1S_0)}{m^6_Q}\times \nonumber \\ &&
   Re\left[\left<^1S_0|\psi^{\dag}\chi|0\right>\left<0|\chi^{\dag}\left(-\frac{i}{2}\overrightarrow{D}\right)^4\psi|^1S_0\right>\right]
\end{eqnarray}

\begin{eqnarray} \label{eq:nq2}
&&\Gamma(^3S_1 \rightarrow
e^+e^-) = \frac{F_{ee}(^3S_1)}{m^2_Q}
\left|\left<0|\chi^{\dag}\sigma\psi|^3S_1\right>\right|^2 \cr
&&+\frac{G_{ee}(^3S_1)}{m^4_Q}
Re\left[\left<^3S_1|\psi^{\dag}\sigma\chi|0\right>\left<0|\chi^{\dag}\sigma\left(-\frac{i}{2}\overrightarrow{D}\right)^2\psi|^3S_1\right>\right]
\cr
&&+\frac{H^1_{ee}\left(^1S_0\right)}{m^6_Q}
\left<^3S_1|\psi^{\dag}\sigma\left(-\frac{i}{2}\overrightarrow{D}\right)^2\chi|0\right> \times \nonumber \\
&&
\left<0|\chi^{\dag}\sigma\left(-\frac{i}{2}\overrightarrow{D}\right)^2\psi|^3S_1\right>+ \frac{H^2_{ee}(^1 S_0)}{m^6_Q}\times \nonumber \\
&&
\ Re \left[\left<^3S_1|\psi^{\dag}\sigma\chi|0\right>
\left<0|\chi^{\dag}\sigma\left(-\frac{i}{2}\overrightarrow{D}\right)^4\psi|^3S_1\right>\right]
\end{eqnarray}

The short distance coefficients F's and G's of the order of $\alpha_s^2$ and $\alpha_s^3$ are given by \cite{bodwin}
\begin{eqnarray}
&&F_{\gamma \gamma}(^1 S_0)=2\pi Q^4 \alpha^2 \left[1+\left(\frac{\pi^2}{4}-5 \right) C_F\frac{\alpha_s}{\pi} \right] \nonumber \\
&&G_{\gamma \gamma}(^1S_0)=-\frac{8 \pi Q^4}{3}\alpha^2 \nonumber \\
&&H^1_{\gamma \gamma}(^1S_0)+H^2_{\gamma\gamma}(^1S_0)=\frac{136\pi}{45} Q^4 \alpha^2
\end{eqnarray}

\begin{eqnarray}
&&G_{ee}(^3 S_1)=-
\frac{8 \pi Q^2}{9} \alpha^2 \nonumber \\
&& H^1_{ee}(^3S_1)+H^2_{ee}(^3S_1)=\frac{58\pi}{54} Q^2 \alpha^2 \nonumber \\
&&F_{ee}(^3S_1)= \frac{2 \pi Q^2 \alpha^2}{3}  \{ 1- 4 C_F
\frac{\alpha_s(m)}{\pi}  +\nonumber \\
&&\left[-117.46+0.82n_f+\frac{140\pi^2}{27} ln\left(\frac{2m}{\mu_A}\right)\right]
(\frac{\alpha_s}{\pi})^2  \}
\end{eqnarray}
The matrix elements that contributes to the decay rates of the S wave states into
 $\eta_Q\rightarrow \gamma \gamma$ and $\psi \rightarrow e^+e^-$ through next-to-leading
 order in $v^2$, the vacuum-saturation approximation gives \cite{GT}\\
\begin{eqnarray}
&&\left<^1S_0|{\cal{O}}(^1S_0)|^1S_0\right>=\left|\left<0|\chi^{\dag}\psi|^1S_0\right>\right|^2[1+
O(v^4 \Gamma)]\nonumber \\
&&\left<^3S_1|{\cal{O}}(^3S_1)|^3S_1\right>=\left|\left<0|\chi^{\dag}\sigma\psi|^3S_1\right>\right|^2[1+O(v^4 \Gamma)]\nonumber \\
&&\left<^1S_0|{\cal{P}}_1(^1S_0)|^1S_0\right>=Re\left[\left<^1S_0|\psi^{\dag}\chi|0\right> \right. \times \nonumber \\
&&\left. \left<0|\chi^{\dag}(-\frac{i}{2}\overrightarrow{D})^2 \psi|^1S_0\right>\right]+ O(v^4\Gamma )\nonumber
\end{eqnarray}

\begin{eqnarray}
&&\left<^3S_1|{\cal{P}}_1(^3S_1)|^3S_1\right>=
Re\left[\left<^3S_1|\psi^{\dag}\sigma\chi|0\right> \right.\times \nonumber\\
&&\left. \left<0|\chi^{\dag}\times \nonumber
\sigma\left(-\frac{i}{2}\overrightarrow{D}\right)^2\psi|^3S_1\right>\right]
+ O(v^4 \Gamma ) \nonumber \\
&&\left<^1S_0|{\cal{Q}}^1_1(^1S_0)|^1S_0\right>=
\left<0|\chi^{\dag}\left(-\frac{i}{2}\overrightarrow{D}\right)^2\psi|^1S_0\right>\nonumber\\
&&\left<^3S_1|{\cal{Q}}^1_1(^3S_1)|^3S_1\right>=\left<0|\chi^{\dag} \sigma
\left(-\frac{i}{2}\overrightarrow{D}\right)^2\psi|^3S_1\right>
\end{eqnarray}

The Vacuum saturation allows the matrix elements of some four fermion operators to be expressed in terms of the regularized wave-function parameters given by \cite{GT}
\begin{eqnarray}
&&\left<^1S_0|{\cal{O}}(^1S_0)|^1S_0\right>=\frac{3}{2 \pi}|R_{P}(0)|^2 \nonumber \\
&&\left<^3S_1|{\cal{O}}(^3S_1)|^3S_1\right>=\frac{3}{2\pi}|R_{V}(0)|^2 \nonumber \\
&&\left<^1S_0|{\cal{P}}_1(^1 S_0)|^1S_0\right>
=-\frac{3}{2 \pi}|\overline{R^*_{P}}\ \overline{\bigtriangledown^2 R_{P}}| \nonumber \\
&&\left<^3S_1|{\cal{P}}_1(^3 S_1)|^3S_1\right>=-\frac{3}{2
\pi}|\overline{R^*_{V}}\ \overline{\bigtriangledown^2 R_{V}}| \nonumber \\
&&\left<^1S_0|{\cal{Q}}^1_1(^1S_0)|^1S_0\right>=
-\sqrt{\frac{3}{2\pi}} \overline{\nabla^2} R_{P}\nonumber \\
&&\left<^3S_1|{\cal{Q}}^1_1(^3S_1)|^3S_1\right>=-
\sqrt{\frac{3}{2\pi}} \overline{ \nabla^2} R_{V} \nonumber \\
\end{eqnarray}

The term $\overline{{\nabla^2} R_{P/V}} $ is the renormalised Laplacian of the radial wave function. We have computed $\overline{{\nabla^2} R_{P/V}} $ term as given by
\cite{khan}.  Accordingly,
\begin{equation}\label{BIND}
 \overline{{\nabla^2} R_{P/V}} =-\epsilon_B R_{P/V} \frac{M_{P/V}}{2}, \ \ \ \ as \ \ \
 r\rightarrow0
\end{equation}
where $\epsilon_B$ is the binding energy and $M$ is the mass of the
respective meson state. The binding energy is computed as
$\epsilon_B=M-(2m_Q)$.\\
The rate of the decay can be estimated in the extreme-nonrelativistic picture, where the system is described by the wave
function for the quark-antiquark pair and depending on their relative position $\vec{r} = \vec{r_c}-\vec{r_{\bar c}}$.
The annihilation takes place at the characteristic distances of order $1/m_c$ which are to be viewed as $r\rightarrow0$
for a nonrelativistic pair, so that the decay amplitude is proportional to the wave function at the origin. So the right
description of meson state through its radial wave function at the origin and its mass along with other model parameters
like $\alpha_s$ and the model quark masses become crucial for the computations of the decay rates. In many cases of potential model predictions, the radial wave functions at the origin are found to overestimate the decay rates. In such cases, it is assumed that the decay of $Q \bar Q$ does not occur at zero separation but at some finite $Q \bar Q$  radial separation. Then arbitrary scaling of the radial wave function at zero separation is done to estimate the decay rates correctly \cite{EE}.\\

In the present study, we have calculated decay properties at zero quark separation ($r=0$) as well as at the finite quark separation $r=r_0$. This radial distance $r_0$ can be considered as the 'colour Compton radius', a quantity related to the electromagnetic processes, as referred by authors in \cite{ajay}. However, particularly in the prediction of the leptonic decay widths considerable improvement has been obtained, when it is evaluated at finite distance $r_0$. The computed di-leptonic decay widths are listed in Table V. The computed di-gamma widths of the $c\bar c$ states are listed in Table VI while the di-gluon widths of S and P wave $c\bar c$ states are listed  in Tables VII and VIII respectively. The computed widths are represented as $\Gamma_{0/0R}(0)$, $\Gamma_{0/0R}(r_0)$ for the di-leptonic and di-gamma decay widths and $\Gamma_{gg/ggR}(0)$, $\Gamma_{gg/ggR}(r_0)$ for the di-gluon decay widths. The computation of di-leptonic and di-gamma decay widths based on NRQCD formalism are listed in Table IX and X respectively.
The quantities, with suffixes carrying R, correspond to the widths with the respective radiative corrections included.

\section{Results and Discussions}
Using the predicted masses and radial wave functions at the origin as well as at finite quark-antiquark separation, the di-gamma, di-leptonic decays of charmonia are computed using conventional Van Royen-Weisskopf non-relativistic formula as well as using NRQCD formalism.
 Apart from this, di-gloun decays of charmonia are also studied using conventional Van Royen-Weisskopf formula .
 The overall agreement of the calculated mass spectra with the experiment \cite{olive} and lattice results \cite{Bali} is impressive. The present study also provides us the importance of the quark mass parameters and the state dependence on the potential strength for the study of the spectral properties of $c\bar c$ mesons. The present study is also an attempt towards the quantitative understanding of the importance of radiative corrections for the decay widths of the heavy flavour quarkonia. The results on the mass spectra of S wave states are shown in the Table I while those for P and D waves with spin-orbit and tensor contributions are shown in the Tables II and III respectively. These results are in good agreement with the available experimental values with just about $1.09\%$ variations, while comparison with those of the lattice QCD predictions  show $1.46\%$ variations.
The precise experimental measurements of the masses of charmonia states provide a real test for the choice of the hyperfine
and the fine structure interactions adopted in the study of charmonia spectroscopy. Hyperfine splitting provides a direct
measure of the strength of the spin-spin chromomagnetic interaction. Recently, charmonium mass splittings in three flavor
lattice QCD has been studied by Fermilab Lattice and MILC collaborations \cite{daniel}. In Table V, we have compared our
results on the mass splittings with the lattice results as well as with the respective experimental results and also with
other potential model predictions.  Both spin-orbit and tensor terms test the strength of the chromoelectric interaction.
The tensor term is in good agreement with lattice as well as experimental results while the spin orbit term is off from the
experimental as well as lattice results. The spin-averaged 1P- 1S splitting tests the central part of the potential. The
splitting of the spin-averaged 2S and 1S levels also tests the â€œcentralâ€ part of the quarkonium effective potential. One of the important feature of the present potential model is that the nature of the quark-antiquark potential is exactly mimic the cornell like potential as seen from Fig 1. Another important feature of the present study is that the decay of charmonia system occurs at a finite range of its separation provided by the color compton radius. This suggests that various processes of quark-antiquark annihilation occur at finite radial separation.

The di-leptonic decay widths computed at finite radial separation defined through the color compton radius are found to be
in better agreement with the experimental values for most of the states. The leptonic decay widths $\Gamma_{0R}$ for 1S state and 3S at finite distance $r_0$ with the inclusion of radiative correction are found to be in good agreement with the experimental data while for 2S state, decay width $\Gamma_{0}$ matches well with experimental results without inclusion of radiative correction.

For 1S and 2S states, computed di-gamma widths $\Gamma_0({r_0})$ at finite quark-antiquark separation without radiative
correction are in good agreement with the experimental results while  for $\chi_{c_0}$ state the results are slightly off
from the experimental results but are in agreement with the other model predictions \cite{Lansberg,Bai}.
The di-gamma decay width $\Gamma_{OR}$-$(r_0)$ predicted for $1^3P_2$ state at finite quark-antiquark separation matches
well with the experimental result, while the decay width $\Gamma_{OR}(0)$ agrees well with the experimental result. Though
we predict di-gamma decay widths of 2P states, they are not measured experimentally. So we have compared our results with
the available other theoretical predictions.

The di-gluon decay widths predicted for the $c\bar c$ system are all in good accord with available experimental data as
well as other model predictions. It is observed that di-gluon decay widths of 1S and 2S states of $c\bar c$ without
radiative corrections and with binding energy effects are consistent with experimental values when evaluated at origin.
On the other side the di-gluon decay widths of 1S and 2S states of $c\bar c$ with radiative corrections and without
inclusion of the binding energy effects are consistent with experimental values when evaluated at some finite distance.

 The predicted di-gluon decay width of the $\chi_{c0}$ state with the inclusion of binding energy effects and without
 radiative corrections agrees well with the experimental values when it is evaluated at origin and finite distance $r_0$.
 For $\chi_{c2}$ state the decay width without inclusion of binding energy effects and without radiative correction is in
 agreement with the experimental value when it is evaluated at the origin. In case of the di-leptonic decay width, RMS
 variation when it is evaluated at finite quark-antiquark separation $r_0$, without and with inclusion of radiative
 corrections is 0.50 and 1.80 respectively which is less than the RMS variation when calculated at origin. So the leptonic
 decay occurs at finite quark-antiquark separation $r_0$.
The RMS variation in di-gamma when evaluated at origin, without and with inclusion of radiative corrections is 4.89 and
2.60 respectively. This RMS variation in di-gamma decay width becomes less when it is evaluated at finite quark-antiquark
separation $r_0$ i.e. it is 3.19 and 1.56, without and with inclusion of radiative corrections respectively.
So in case of di-gamma decay , finite separation as well as radiative corrections both are important.
There is a large RMS variation in the di-gluon decay width when it is calculated with quark mass m. But this variation
decreases when it is evaluated with the inclusion of binding energy effects (i.e. with effective quark mass).
In case of  di-gluon decay width, the RMS variation is 2.46 and 6.55 when evaluated at zero quark-antiquark separation
without and with inclusion of radiative corrections. But the RMS variation in di-gluon decay width is 11.2 and 9.61 when
evaluated at $r_0$ without and with inclusion of radiative corrections. So in case of di-gluon decay finite separation is
found not important. We predict the di-gluon decay width of 3S and 2P states of charmonia and we look forward to see the experimental support in favour of our predictions.
In the NRQCD formalism the di-leptonic and di-gamma decay widths have been computed by using the radial wavefunctions and their derivatives at origin as well at some finite distance seperation. The predicted di-leptonic decay widths evaluated at origin with and without inclusion of radiative corrections are found to be overestimated while those who are evaluated at some finite separation are found to be in better agreement with the experimental data as well as other theoretical predictions. The same trend is seen in the case of the di-gamma decay widths. With NRQCD formalism, the RMS variation in the di-leptonic and di-gamma decay are 0.29 and 0.83 respectively when evaluated at finite radial separation. It can be concluded that NRQCD formalism has most of the corrective contributions required for most of the potential models for the right predictions of the decay rates. Finally, we believe that future high luminosity experiments will be able to shed more light in the understanding of the
quark-antiquark interaction.

\section{Acknowledgments}
The work is part of Major research project NO. F. 40-457/2011(SR) funded by UGC, INDIA.

\end{multicols}

\begin{multicols}{2}

\end{multicols}

\clearpage

\end{document}